# Plasmonic quasicrystals for designable ultra broadband transmission enhancement and second harmonic generation

Sachin Kasture, Ajith P R, V J Yallapragada, Raj Patil, Nikesh V. V., Gajendra Mulay, Achanta Venu Gopal*

DCMPMS, Tata Institute of Fundamental Research, Homi Bhabha Road, Mumbai 400 005

achanta@tifr.res.in

Quasi-crystals are intriguing as they exhibit rotational symmetry and long range ordering but lack translational symmetry. 2-dimensional metal-dielectric patterns are interesting to make use of surface plasmon polariton (SPP) mediated local field enhancement and for near dispersionless SPP modes. In plasmonic crystals, the orientation and periodicity of the pattern dictate the polarization response and the discrete plasmon resonances while the interfaces define the plasmon dispersion. However, unique properties of plasmonic quasicrystals lead to polarization independence, designable k-space and broadband transmission enhancement due to SPP mediation. These are useful in many applications like energy harvesting, nonlinear optics and quantum plasmonics. We demonstrate design and fabrication of large area quasicrystal air hole patterns of $\pi/5$ symmetry in metal film in which broadband, launch angle and polarization independent transmission enhancement as well as broadband second harmonic generation are observed. Designable transmission response, other symmetries and tilings are possible.

Sub-wavelength patterns in the form of grooves, hole arrays, or corrugations on metal-dielectric interfaces help optically excite surface plasmon polaritons (SPPs) at the interface[1]. SPPs enhance transmission as well as local field which have wide ranging applications in nanophotonics[2-12]. Plasmonic crystals are periodic metal-dielectric structures realized for a specific application in which the SPPs are optically excited by periodic structures[1]. The translation symmetry of these crystals in a given direction results in discrete SPP excitation as well as in polarization dependence. For example, the longitudinal SPP modes are excited by 1-dimensional patterns (gratings) by the transverse magnetic (TM) field though under specific orientations transverse electric field may excite SPPs[13]. In general, when light of wave vector $k_0$ diffracted by a 3-dimensional periodic pattern excites SPPs, the momentum conservation is given by $k_x^2+k_y^2+k_z^2 = k_{SPP}^2$, where $k_x = k_0\sin(\theta)\cos(\varphi) \pm i2\pi/a_x$, $k_y = k_0\sin(\theta)\sin(\varphi) \pm j2\pi/a_y$ and $k_z = k_0\cos(\theta) \pm k2\pi/a_z$ in which i, j, k are integers, $a_x$, $a_y$, $a_z$ are lattice constants in the 3 directions, $\theta$ is the angle of incidence and $\varphi$ is the azimuthal angle. Thus, the orientation of the sample or the azimuthal angle $\varphi$ and thus polarization of light defines the SPP mode excited. Unlike crystals, quasicrystals have only long range ordering and rotational symmetry[14]. In quasi crystal patterns, discrete SPP modes[15], extra-ordinary transmission in different wavelength regions[16-22], possibility to design for specific phase matching conditions for harmonic generation[23,24], focusing and directivity of emission[25,26] have been demonstrated. In addition to the designable k-space, broadband local field and transmission enhancement in plasmonic quasicrystals (PQCs) would be useful. For example, for light harvesting, the top (or bottom) contact of the photovoltaic device can be patterned with quasicrystal structure, in nonlinear optics for harmonic generation (SHG) though the metal surface optical nonlinearity is weak ($\chi^{(2)}$)[27], in quantum plasmonics for high quality factor cavities[28]. Though in random patterns one can achieve broadband response[29], it is in practice not easy to design them for specific k-space response. Aperiodic structures and their applications to nanophotonics are reported earlier[30,31].

In order to design quasicrystal structures several methods based on tiling, grid, inflation/deflation, projection, decoration, cluster model, etc., are employed[30,32]. In most of the previous reports on plasmonic quasicrystals, the dual grid method or the Penrose tiling using fat and thin rhombii tiles for designing and ion beam milling for patterning the thin metal film were employed[16-26]. Electron beam lithography has been used for making gold nanodots arranged in aperiodic patterns[29] and Moire nanolithography has been used to demonstrate quasicrystal patterns of high rotational symmetry[33]. In optical domain, the effect of size of the pattern on the transmission enhancement was studied as well as the origin of transmission enhancement was proposed to be not related to long range ordering[16,17]. Both these studies are on patterns covering an area of 10 x 10 $\mu m^2$ milled in silver thin films and thus are comparable to or smaller than the propagation length of SPPs. We present a method to realize quasicrystal patterns over large area using electron beam lithography. The design and fabrication methods can be easily extended to any rotational symmetry and Penrose tiles. We present in PQCs with π/5 symmetry, polarization independent, near dispersionless broadband transmission enhancement compared to unpatterned metal. Plasmon mediated local field enhancement has also been used to demonstrate second harmonic generation over the broad band. We realize ultra broadband transmission enhancement demonstrating PQCs with designable spectral response.

In oblique tiling method, a set of parallel lines tilted by π/n angle are superimposed by n similar sets of parallel lines each rotated by π/n with respect to the previous set. The intersection points of the n sets of these parallel lines give the coordinates of the quasicrystal with n-fold symmetry. To design the structure, we replace the continuous lines with dots and thus the set of parallel lines are now a 2-D array of dots as shown in Figure 1a. By successively superimposing *n* such 2-dimensional arrays of dots that are rotated by π/n with respect to the previous array, we generate a hole pattern. The spatial points which are common to all the n overlays give the coordinates of the QC of π/n rotational symmetry. The intrinsic period of the starting 2-d array defines the center of the spectral response and thus can be tuned to any wavelength required. The complexity of the problem is in identifying the coordinates of the quasicrystal lattice points from the large set of points for electron beam lithography. To find repetitions in the large array of points, we used QuickSort algorithm which is a sort-in-place algorithm that does not need additional memory and sorts an array of size N in O(N × log(N)) time[34]. This helps identify repetitions in approximately O(N) time which is much faster than the time taken by brute force methods (O($N^2$)). In addition, we can scale the pattern for specific k-response by using the scaling property of Fourier space that is for a constant *c* and real and phase space functions given by F and f,

$$\int_{-\infty}^{\infty} F(ck)e^{ikx}dk = \frac{1}{c}f(\frac{x}{c}) \qquad (1)$$

This method was used to generate the coordinates over a large area of ~ 1mm$^2$. By electron beam lithography and dry etching, we patterned thin gold film on optically flat quartz (see Methods). Fig.1b shows the SEM image of the fabricated quasicrystal. The initial rhombus used in designing the structure had 1mm long sides each having holes at a periodic separation (*a* in Fig.1a) of 600 nm. That is, there are about 2.8 million points in 1 mm$^2$ area and with 5 superimpositions there are a total of about 14 million points. Of these, the lattice points corresponding to QC of π/5 rotational symmetry are about 12 million points.

In the following we use PQCs in which all the holes excepting those that are within 10nm vicinity of a given hole are retained to avoid over exposure during e-beam lithography.

This sorting results in, the coordinates of both QC of π/5 symmetry as well as those corresponding to the 5 base lattices, about 13 million holes in 1 mm² area. The average hole diameter obtained was 92 ± 7 nm by careful e-beam dosage. The diffraction pattern (Fig.1c) recorded using a 635nm diode laser beam clearly shows π/5 symmetry. The observed multiple concentric circles show that the structure displays ordering at different levels. Fig. 1d shows the calculated Fourier transform of the structure shown in Fig.1a.

We performed angle resolved white light transmission measurements to measure the normalized transmission (ratio of transmission through paterned and unpatterned Gold) for TM, and TE and unpolarized light for differnt θ. Fig. 2a shows the spectra for θ = 5° and φ = 0°. We observe upto 9 fold transmission enhancement with respect to the unpatterned metal tranmission and near dispersionless broadband response (550 – 720 nm).

To calculate the SPP resonances and their dispersion, for a periodic pattern, there are theoretical models for a given interface[35,1]. Analytical expressions for SPP dispersion for two layers (single interface) and three layers (two interfaces) are well known[35]. In order to calculate the grating excited SPP modes one can equate the 2-d momentum conservation equation where appropriate $k_{SPP}$ is used depending on the layer structure. For example, for 2-d pattern of air holes[36],

$$\sqrt{\left(k_0 \sin(\theta)\cos(\varphi) \pm \frac{m2\pi}{a_1}\right)^2 + \left(k_0 \sin(\theta)\sin(\varphi) \pm \frac{n2\pi}{a_2}\right)^2} = k_{SPP} \quad (2)$$

where $a_1$ and $a_2$ are the periods in x and y directions, respectively. Compared to a periodic crystal, the azimuthal angle (φ) dependence in a quasicrystal is negligible due to rotational symmetry. In addition, due to lack of translational symmetry, the Bloch periodic terms ($m2\pi/a_1$ and $n2\pi/a_2$) do not exist in the quasi crystals and are to be replaced with the reciprocal vectors which can be obtained from either the calculated or measured diffraction patterns. Thus, the quasi momentum conservation rule for QCs is given by[15], $k_{//} + k_{SPP} = F^{(i)}$ where $k_{//}$ and $k_{SPP}$ are the in-plane components of the wave vector of incident light and SPP, respectively and $F^{(i)}$ are the reciprocal vectors. From the measured diffraction pattern, one can extract the $F^{(i)}$s and thus calculate the $k_{SPP}$s.

We use the Fourier transform (FT) shown in Fig.1d to deduce the reciprocal vectors ($k_I$, $k_{II}$ etc.,) corresponding to the successive circles. The values are $k_I$ = 1.82μm⁻¹ (red) and $k_{II}$ = 2.91μm⁻¹ (green). Each SPP mode may be denoted in terms of a tuple (K, B, m, n). K takes the values $k_I$, $k_{II}$, B takes the values Q (quartz) or A (air) which defines the interface with gold, over which the SPP propagates and (m,n) are integers. The calculated positions of various modes are shown in the measured SPP dispersion contour plot (data not shown). For the calculations, the values of $\varepsilon_m$ are taken from Johnson & Christy parameters[37] for gold and $\varepsilon_d$ = 1, 2.25 for air and quartz, respectively. The resonance corresponding to $k_{II}$ is more prominent as can be seen in all the plots. This is consistent with the stronger diffraction maxima in Fig. 1d for $K_{II}$ compared to $K_I$. The resonance ($k_I$,A,0,1), is a near dispersionless mode[36] for φ = 0°. Figure 2a shows the measurement geometry.

From Fig.2b, we can also see that the transmission spectra are similar for TE, TM and unpolarized light. Broad transmission enhancement is seen in the spectra for all polarizations at θ = 5°. Polarization independence is a manifestation of rotational symmetry with no translational symmetry[38]. In addition, the multiple periodicities displayed lead to greater number of SPP modes in the same wavelength range compared to 1-d or 2-d periodic structures, resulting in broad band transmission enhancement. To make sure that the observed enhancement is not because of the high hole density, we tested a structure with 2-d square array of holes in metal with same metal thickness and with the same hole density as the quasicrystal sample. We observed much lower enhancements (~1.1 compared to unpatterned metal) for this structure (data not presented). We also checked the dependence on the hole size distribution by studying pattern with larger hole size distribution between 115 ± 35 nm. We found almost identical transmission behaviour.

In general, the 2ⁿᵈ order optical susceptibility related to surface nonlinearity of gold is weak to observe SHG. But the plasmon mediated local field enhancement increases the 2$^{nd}$ order polarization resulting in SHG. Due to polarization independent and near dispersionless transmission enhancement, we could get broadband measurable SHG. Figure 3a shows the measured quadratic dependence of the SHG power on the fundamental power at 800nm. Figure 3b shows the SHG signal measured at different wavelengths by tuning the fundamental wavelength. Unlike in crystals, here the SHG is seen even for normal incidence of fundamental. The measured SHG is in the direction of the fundamental and we did not find θ and φ dependence. The gold surface $\chi^{(2)}$ at different wavelengths is calculated based on method presented in Ref.39. Fig. 3c shows the estimated second order optical susceptibility at different wavelengths from the measured linear and nonlinear optical measurements.

It is advantageous to have control over the design to realize QC structures with desired spectral response. This can have applications in light harvesting, filters, nonlinear optics and quantum plasmonics where manipulation of dipole emitters is needed. We show such feasibility in a dual period QC realized by combining two base periods of 0.6 and 1 µm in which the transmission band has been increased to 800 nm. Fig. 4 shows more than an order of magnitude enhancement in transmission over a broad wavelength range. Thus, we can design and realize plasmonic quasicrystals of any required symmetry and transmission response.

**Conclusion**

We demonstrated a method to design and fabricate large area quasi-crystal patterns of a given rotational symmetry by electron beam lithography. As an example, we demonstrated a π/5 plasmonic quasicrystal with its clear optical diffraction pattern. These structures showed near dispersionless and polarization independent enhanced transmission over a broad wavelength range. We also showed launch angle and polarization independent second harmonic generation over the broad wavelength range due to enhanced local field at the interfaces due to SPP excitation. The scalability of the Fourier transform of the designed structure helps tune the quasicrystal pattern for a specific wavelength without designing each time. Also, by combining different base periods one may design the k-space response required. We demonstrated one such structure with very broadband transmission enhancement of over an order of magnitude.

**References**

1. Raether, H. *Surface Plasmons on Smooth and Rough Surfaces and on Gratings* (Springer, 1986).
2. Ebbesen, T. W., Lezec, H. J., Ghaemi, H. F., Thio, T. & Wolff, P. A. Extraordinary optical transmission through sub-wavelength hole arrays. *Nature* **391**, 667-669 (1998).
3. Martín-Moreno, L., et al. T. W. Theory of Extraordinary Optical Transmission through Subwavelength Hole Arrays. *Phys. Rev. Lett.* **86**, 1114-1117 (2001).
4. Belotelov, V. I., et al. Enhanced magneto-optical effects in magnetoplasmonic crystals. *Nature NanoTech.* **6**, 370-376 (2011).
5. Belotelov, V. I., et al. Plasmon mediated magneto-optical transparency. *Nature Commun.* **4**, 2128 (2013).
6. Zhang, J. Z. & Noguez, C. Plasmonic optical properties and applications of metal nanostructures. *Plasmonics.* **3**, 127-150 (2008).
7. Ma, R-M., Oulton, R. F., Sorger, V. J., Bartal, G. & Zhang, X. Room-temperature sub-diffraction-limited plasmon laser by total internal reflection. *Nature Materials.* **10**, 110-113 (2011).
8. Tvingstedt, K., Persson, N-K., Inganäs, O., Rahachou, A. & Zozoulenko, I. V. Surface plasmon increase absorption in polymer photovoltaic cells. *Appl. Phys. Lett.* **91**, 113514 (2007).
9. Barnes, W. L., Dereux, A. & Ebbesen, T. W. Surface plasmon subwavelength optics. *Nature* **424**, 824-830 (2003).
10. Righini, M., Zelenina, A. S., Girard, C. & Quidant, R. Parallel and selective trapping in a patterned plasmonic landscape. *Nat. Phys.* **3**, 477-480 (2007).
11. Zhang, X., Liu, H., Tian, J., Song, Y. & Wang, L. Band-selective optical polarizer based on gold-nanowire plasmonic diffraction gratings, *Nano Lett.* **8**, 2653-2658 (2008).

12. Antoine, R., et al. Surface plasmon enhanced second harmonic response from gold clusters embedded in an alumina matrix. *J. Appl. Phys.* **84**, 4532-4536 (1998).
13. Vengurlekar, A. S. Polarization dependence of optical properties of metallodielectric gratings with subwavelength grooves in classical and conical mounts. *J. Appl. Phys.* **104**, 023109 (2008).
14. Penrose, The role of aesthetics in pure and applied mathematical research. R. *Bull Inst. Math. Its Appl.* **10**, 266-271 (1974).
15. Matsui, T., Agrawal, A., Nahata, A. & Vardeny, Z. V., Transmission resonances through aperiodic arrays of subwavelength apertures. *Nature* **446**, 517-521 (2007).
16. Przybilla, F., Genet, C. & Ebbesen, T. W., Enhanced transmission through Penrose subwavelength hole arrays. *Appl. Phys. Lett.* **89,** 121115 (2006).
17. Pacifici, D., Lezec, H. J., Sweatlock, L. A., Walters, R. J. & Atwater, H. A. Universal optical transmission features in periodic and quasiperiodic hole arrays. *Opt. Express.* **16,** 9222-9238 (2008).
18. Papasimakis, N., Fedotov, V. A., Schwanecke, A. S., Zheludev, N. I. & Garci´a de Abajo, F. J. Enhanced microwave transmission through quasicrystal hole arrays. *Appl. Phys. Lett.* **91,** 081503 (2007).
19. Hao, R., Jia, H., Ye, Y., Liu, F., Qiu, C., Ke, M. & Liu, Z. Exotic acoustic transmission through hard plates perforated with quasiperiodic subwavelength apertures. *Europhys. Lett.* **92,** 24006 (2010).
20. Agrawal, A., Matsui, T., Vardeny, S. V. & Nahata, A. Terahertz transmission properties of quasiperiodic and aperiodic aperture arrays. *J. Opt. Soc. Am. B.* **24,** 2545-2555 (2007).
21. Rockstuhl, C., Lederer, F., Zentgraf, T. & Giessen, H. Enhanced transmission of periodic, quasiperiodic, and random nanoaperture arrays, *Appl. Phys. Lett.* **91**, 151109 (2007).
22. Xue, J., Zhou, W-Z., Dong, B-Q., Wang, X., Chen, Y., Huq, E., Zeng, W., Qu, X-P. & Liu, R. Surface plasmon enhanced transmission through planar gold quasicrystals fabricated by focused ion beam technique. *Microelectron. Eng.* **86**, 1131-1133 (2009).
23. Lifshitz, R., Arie, A. & Bahabad, A. Photonic quasicrystals for nonlinear optical frequency conversion. *Phys. Rev. Lett.* **95**, 133901 (2005).
24. Xu, T., Zhang, G. P. & Blair, S. Second harmonic emission from sub-wavelength apertures: Effects of aperture symmetry and lattice arrangement. *Opt. Expr.* **15**, 13894-13906 (2007).
25. Huang, F. M., Zheludev, N., chen, Y. & Garcia de Abajo, F. J. Focusing of light by a nanohole array. *Appl. Phys. Lett.* **90**, 091119 (2007).
26. Regan, C. J., Grave de Peralta, L. & Bernussi, A. A. Directivity and isotropic band-gap in 12-fold symmetry plasmonic quasicrystals with small index contrast. *Appl. Phys. Lett.* 99, 181104 (2011).
27. Jha, S. S. Theory of optical harmonic generation at a metal surface. *Phys. Rev.* **140**, A2020-A2030 (1965).
28. Min, B., Ostby, E., Sorger, V., Ulin-Avila, E., Yang, L., Zhang, X. & Vahala, K. High-Q surface-plasmon-polariton whispering gallery microcavity. *Nature* **457**, 455-458 (2009).
29. Gopinath, A., Boriskina, S. V., Feng, N-N., Reinhard, B. M. & Negro, L. D. Photonic-plasmonic scattering resonances in deterministic aperiodic structures. *Nano Lett.* **8**, 2423-2431 (2008).
30. Macia, E. Exploting aperiodic designs in nanophotonic devices. *Rep. Prog. Phys.* **75**, 036502 (2012).
31. Vardeny, Z. V., Nahata, A. & Agrawal, A. Optics of photonic quasicrystals. *Nature Photonics* **7**, 177-187 (2013).
32. Socolar, J. E. S., Steinhardt, P. J. & Levine, D. Quasicrystals with arbitrary orientational symmetry. *Phys. Rev. B.* **32**, 5547-5550 (1985).
33. Lubin, S. M., Zhou, W., Hryn, A. J., Huntington, M. D. & Odom, T. W. High rotational symmetry lattices fabricated by Moire nanolithography. *Nano Lett.* **12**, 4948-4952 (2012).
34. Cormen, T. H., Leiserson, C. E., Rivest, R. L. & Stein, C. *Introduction to Algorithms*

(MIT Press, McGraw Hill Book Company, 2001).
35. Maier, S. A. *Plasmonics: Fundamentals and Applications*, (Springer, 2007).
36. Kasture, S., et al. *Near dispersion-less surface plasmon polariton resonances at a metal-dielectric interface with patterned dielectric on top. Appl. Phys. Lett*. **101**, 091602 (2012).
37. Johnson P. B. & Christy, R.W. Optical constants of the noble metals. *Phys. Rev. B*. **6**, 4370-4379 (1972).
38. Bliokh, Y. P., Brodsky, Y. L., Chashka, Kh. B., Felsteiner, J. & Slutsker, Y. Z. Broad-band polarization-independent absorption of electromagnetic waves by an overdense plasma. *Phys. Plasmas.* **17**, 083302 (2010).
39. Dick, B., Gierulski, A. & Marowsky, G. Determination of the nonlinear optical susceptibility $\chi^{(2)}$ of surface layers by sum and difference frequency generation in reflection and transmission. *Appl. Phys. B.* **38**, 107-116 (1985).

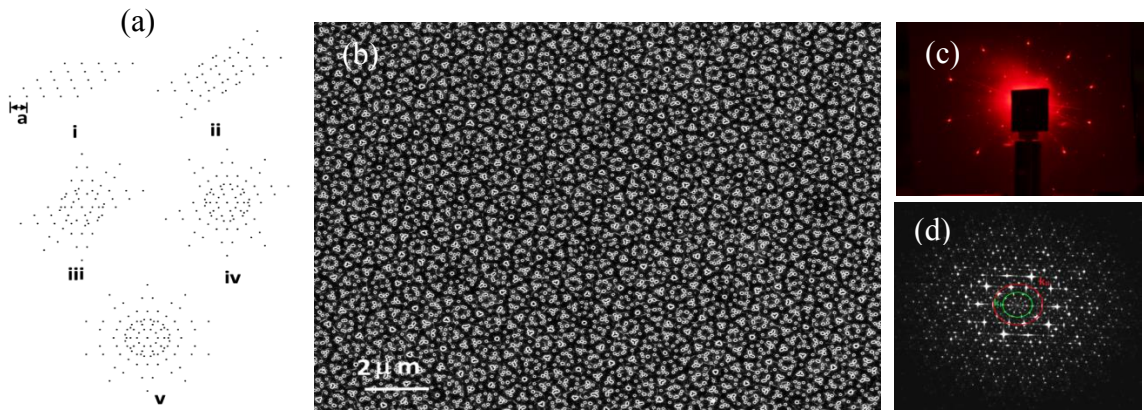

**Fig. 1** (a) Schematic showing the steps in designing of quasicrystal with π/5 rotational symmetry. (b) SEM image of the fabricated quasicrystal pattern. (c) Measured diffraction pattern of the quasi crystal structure for 635nm light. (d) Calculated Fourier transform of the final pattern shown in (a).

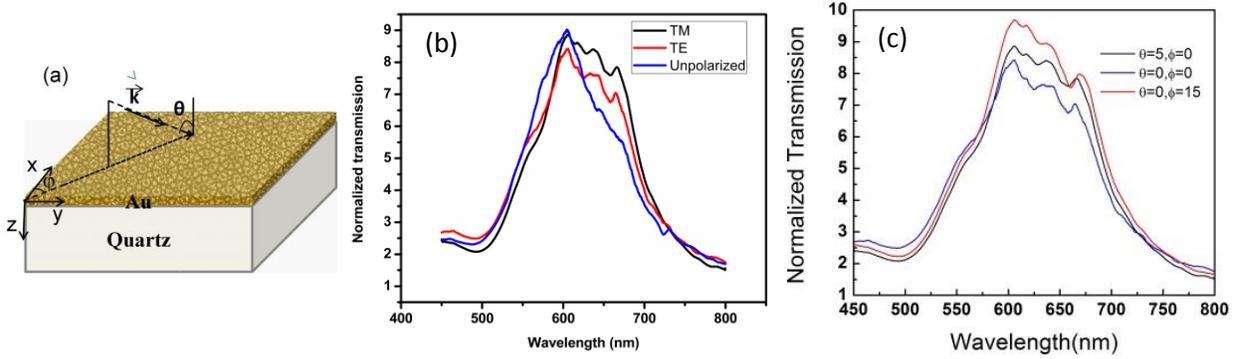

Figure 2 Polarization and angle independence of the transmission enhancement (a) Schematic of the measurement geometry. Normalized transmission spectra are shown for different incident polarizations (b) and for different θ and ϕ for TM polarized light (c).

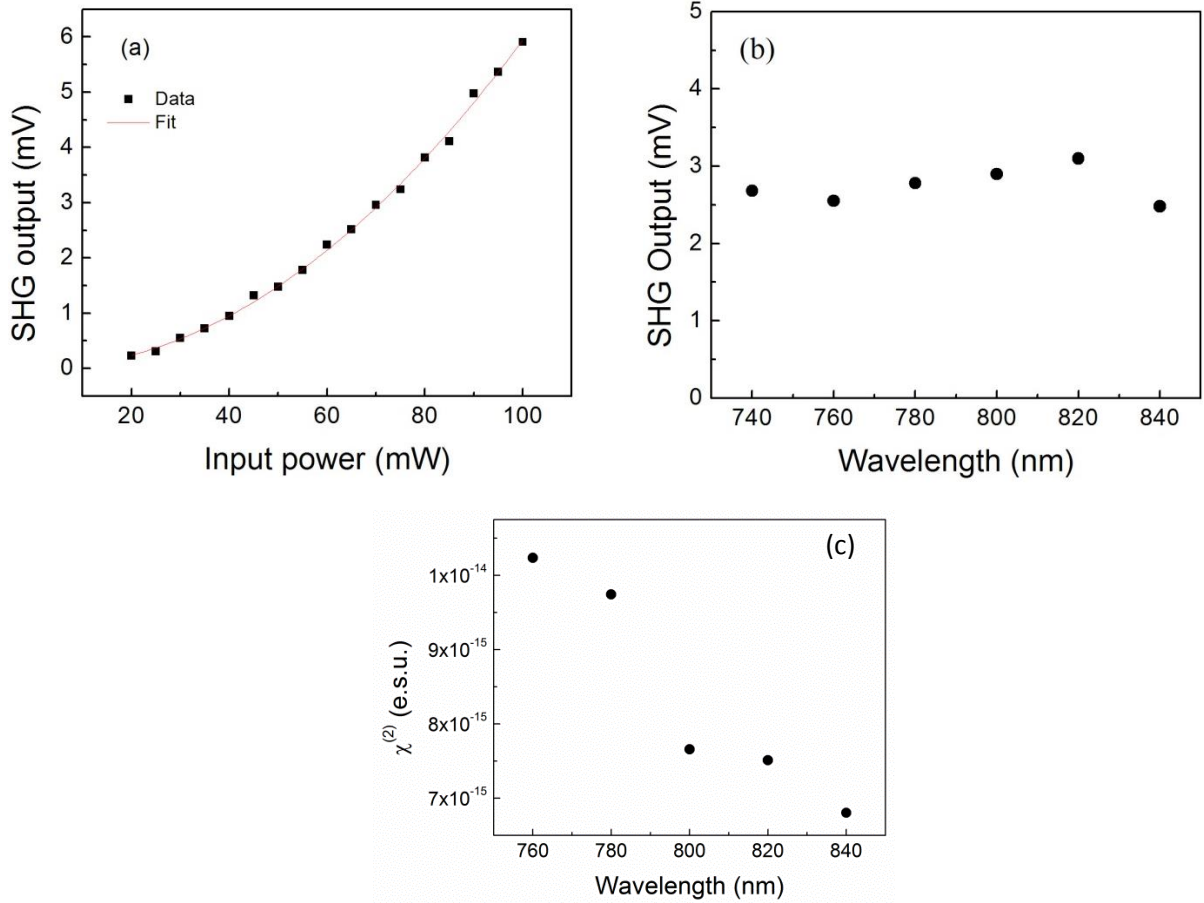

**Figure 3** (a) Shows the measured quadratic power dependence of SHG power on the incident power for 800nm fundamental. Squares are measured points and line is a quadratic fit. (b) shows the measured SHG power generated over the broadband due to SPP mediated local field enhancement of the fundamental beam at the interface. (c) shows the estimated $\chi^{(2)}$ from the linear and nonlinear optical measurements.

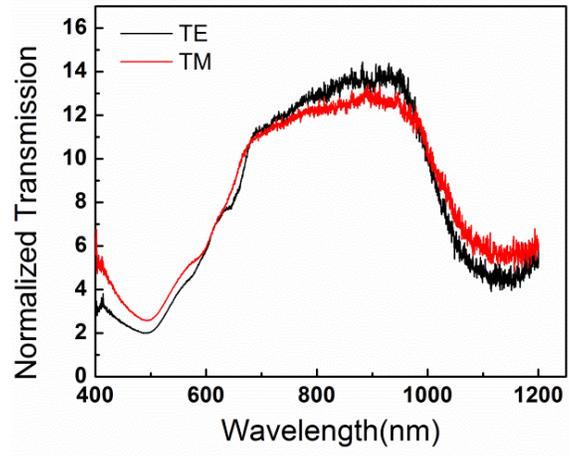

**Fig. 4** Very broadband enhanced transmission in bi-period quasicrsytal structure compared to unpatterned gold and its polarization independence.